# Ubiquitous Superconducting Diode Effect in Superconductor Thin Films

Yasen Hou[1*], Fabrizio Nichele[2], Hang Chi[1,3], Alessandro Lodesani[1], Yingying Wu[1], Markus F. Ritter[2],

Daniel Z. Haxell[2], Margarita Davydova[4], Stefan Ilić[5], Ourania Glezakou-Elbert[6], Amith Varambally[7], F.

Sebastian Bergeret[5,8], Akashdeep Kamra[9*], Liang Fu[4], Patrick A. Lee[4*], Jagadeesh S. Moodera[1,4*]

[1]*Francis Bitter Magnet Laboratory and Plasma Science and Fusion Center, Massachusetts Institute of
Technology, Cambridge, MA 02139, USA*

[2]*IBM Research Europe - Zurich, Säumerstrasse 4, 8803 Rüschlikon, Switzerland*

[3]*U.S. Army CCDC Army Research Laboratory, Adelphi, Maryland 20783, USA*

[4]*Department of Physics, Massachusetts Institute of Technology, Cambridge, MA 02139, USA*

[5]*Centro de Física de Materiales (CFM-MPC), Centro Mixto CSIC-UPV/EHU, Pº Manuel de
Lardizabal 5, Donostia-San Sebastián 20018, Spain*

[6]*Hanford High School, Richland, Washington 99354, USA*

[7]*Vestavia Hills High School, Vestavia Hills, Alabama 35216, USA*

[8]*Donostia International Physics Center (DIPC), Donostia-San Sebastián 20018, Spain*

[9]*Condensed Matter Physics Center (IFIMAC) and Departamento de Física Teórica de la Materia
Condensada, Universidad Autónoma de Madrid, E-28049 Madrid, Spain*

* Corresponding authors. Email: yshou@mit.edu (Yasen Hou); akashdeep.kamra@uam.es (Akashdeep
Kamra); palee@mit.edu (Patrick A. Lee); moodera@mit.edu (Jagadeesh S. Moodera)

**ABSTRACT**

   The macroscopic coherence in superconductors supports dissipationless supercurrents which could play

a central role in emerging quantum technologies. Accomplishing unequal supercurrents in the forward and

backward directions would enable unprecedented functionalities. This nonreciprocity of critical

supercurrents is called superconducting (SC) diode effect. We demonstrate strong SC diode effect in

conventional SC thin films, such as niobium and vanadium, employing external magnetic fields as small as

1 Oe. Interfacing the SC layer with a ferromagnetic semiconductor EuS, we further accomplish non-volatile

SC diode effect reaching a giant efficiency of 65%. By careful control experiments and theoretical modeling, we demonstrate that the critical supercurrent nonreciprocity in SC thin films could be easily accomplished with asymmetrical vortex edge/surface barriers and the universal Meissner screening current governing the critical currents. Our engineering of the SC diode effect in simple systems opens door for novel technologies. Meanwhile, we reveal the ubiquity of Meissner screening effect induced SC diode effect in superconducting films, which should be eliminated with great care in the search of exotic superconducting states harboring finite-momentum Cooper pairing.

**Introduction**

Similar to a conventional semiconductor diode, a superconductor with non-reciprocal current flow, a SC diode, may form the building block for, e.g., dissipationless SC digital logic. The recent observation of such a SC diode effect in a complex thin film superconductor heterostructure subjected to an external magnetic field has stimulated vigorous activity towards understanding and replicating it [1]. Supercurrent rectification has also been demonstrated in multiple Josephson junction systems including Al-InGaAs/InAs-Al [2], NbSe$_2$/Nb$_3$Br$_8$/NbSe$_2$ [3] and Nb-NiTe$_2$-Nb [4], where largest nonreciprocities are observed at large in-plane magnetic fields [2,4]. Furthermore, an intrinsic SC diode effect has been observed in few-layer NbSe$_2$ [5] and twisted trilayer graphene/WTe$_2$ heterostructures [6] in an out-of-plane magnetic field. To quantify the diode effect, it is common to introduce an asymmetry parameter, called the diode efficiency, $\eta = \frac{I_c^+ - I_c^-}{I_c^+ + I_c^-}$ where $I_c^+$ and $I_c^-$ are the critical currents in the two opposite directions. The value of $\eta$ denotes the magnitude of the diode effect, while the sign defines the polarity. Up to now, reported values of $\eta$ range from a few percent to 30% [1-6].

Several theoretical mechanisms have been proposed to explain the SC diode effects in superconductors [7-11] and in Josephson junctions [12,13], with special emphasis on the potential role of the finite-momentum Cooper pairing [7-13]. While this mechanism focuses on the intrinsic depairing current [7-13], it is known that nearly all the superconductor films fail to be governed by the critical pair breaking mechanism, which merely offers the theoretical maximum for a specially designed sample [14-16]. A broad range of other mechanisms come to govern different samples [17,18]. Magnetic flux, or Abrikosov vortices, are normally pinned to defect centers or surfaces of a superconductor [19]. The current flow, however, produces a Lorentz force on the vortices. A critical current is often measured when the pinning centers or the surface barrier cannot hold vortices anymore and dissipation starts in the superconductor [18,20-22]. This principle has been exploited to engineer superconducting vortices-based rectifiers [23-27]. Furthermore, Vodolazov and Peeters predicted existence and engineering of SC diode effect employing controlled edge disorder [23]. This escaped experimental realization thus far and the present work accomplishes it.

Here, using V and Nb superconductors, three types of the SC diode effect are demonstrated, two of which are rooted in the universal Meissner screening instead of the rare finite-momentum Cooper pairing. We show a robust control of the nonreciprocity with record high efficiency in conventional superconductors without requiring additional spin-orbit coupling (SOC), exchange fields.

## Results

The V, Nb and EuS films in our experiments were deposited on clean, heated sapphire substrates in a single deposition process in a molecular-beam epitaxy chamber (base pressure $< 4 \times 10^{-10}$ Torr) [28]. The thicknesses ($d_m$, $d$) of EuS and V (or Nb) were respectively 5 and 8 nm. The film was patterned by e-beam lithography and Ar ion milling into a Hall bar geometry, with width $W \sim 8$ μm and length $L \sim 48$ μm [Fig. 1(a)]. Patterned V device had $T_c$ 3.5 K $\sim$ 4.3 K and a residual resistance ratio around 3 (Supplemental Material [29]). Four-probe geometry current vs voltage scans were measured to observe the critical current ($I_c$) nonreciprocity. The $I$-$V$ scans taken at 1.8 K for a V film subjected to out-of-plane magnetic field are shown in Fig. 1(b). Increasing and decreasing current sweeps showed distinctly different $I_c$ values: $I_c^{\pm}$ marking the SC to normal state transition, whereas a much smaller retrapping current $I_r^{\pm}$ was recorded while transitioning back from the normal to the SC state [Fig. 1(b)]. The low $I_r^{\pm}$ is often attributed to self-heating when the film is in the normal state [3,4]. In this study, we focus on $I_c^{\pm}$ where the nonreciprocity was controllable by applying an out-of-plane magnetic field. With 2.8 Oe field applied along the $+z$ direction, $I_c$ for positive direction ($I_c^+$) was significantly larger than when the current flow was in the negative direction ($I_c^-$). By reversing the magnetic field $I_c^+$ and $I_c^-$ magnitudes interchanged. The magnetic field dependence of $I_c^+$ and $I_c^-$ is plotted in Fig. 1(c), $I_c$ showing large field dependence, an "inverted V" shape, with peaks occurring at ±2.5 Oe. A noticeable current rectification occurred even for fields <1 Oe, its polarity controllable by the field direction. The diode efficiency versus magnetic field, Fig. 1(c) exhibiting a maximum efficiency of ~19% at ±2.8 Oe. Such supercurrent rectification (type A) was observed in all the superconducting devices we measured. $I$-$V$ scans of another V device and a Nb device are presented in Supplemental Material [29], showing similar $I_c$ nonreciprocity for an out-of-plane magnetic field.

For the current flow in a SC without breaking the mirror symmetry with respect to the $x$-$z$ plane, the $+x$ and $-x$ directions are equivalent. Thus, we attribute the observed diode effect to a combination of Meissner current generated to screen the applied magnetic field and symmetry breaking of the device edges during fabrication. In practice, the two edges of a SC stripe could never be identical, thereby admitting slightly different critical current densities $j_c$ and $j_c + \delta j_c$, which are smaller than the Ginzburg-Landau depairing limit $j_{GL}$, as indicated in Fig. 1(d). When current density in the device is above $j_c$, the Lorentz force on vortices nucleated at the edge overcomes the Bean-Livingston barrier thereby enabling vortex flow though the sample which destroys superconductivity [20,23,30-32]. The Meissner effect induces two dissipationless counter-flowing screening current densities $\pm j(B_z)$ at the two edges, when an out-of-plane field is applied. This screening current adds or subtracts to the applied current at opposite edges, and modifies the measured values of $j_c^+$ and $j_c^-$. At small fields, the screening current density is simply the Meissner response $j(B_z) = a B_z$, linear with the applied magnetic field, where $a$ is a constant. For the case shown in Fig. 1(d), on the $B_z > 0$ side,

$$j_c^+ = j_c + a B_z, \quad j_c^- = j_c - a B_z, \; when \; a B_z < \frac{\delta j_c}{2} \tag{1}$$

$$j_c^+ = j_c + \delta j_c - a B_z, \quad j_c^- = j_c - a B_z, \; when \; a B_z \geq \frac{\delta j_c}{2} \tag{2}$$

Similar results are obtained for reversed field. Assuming a uniform current flow in the superconducting stipe, we have $I_c^{\pm} = Sj_c^{\pm}$, where $S$ is the device cross-section.

$$I_c^+ = S(j_c + aB_z), \quad I_c^- = S(j_c - aB_z), \quad when \ aB_z < \frac{\delta j_c}{2} \tag{3}$$

$$I_c^+ = S(j_c + \delta j_c - aB_z), \quad I_c^- = S(j_c - aB_z), \quad when \ aB_z \geq \frac{\delta j_c}{2} \tag{4}$$

The experimental data could be fitted with $Sj_c = 3.14 \ mA, S\delta j_c = 1.35 \ mA$ and $Sa = 0.275 \ mA.Oe^{-1}$ as shown by the dashed curves in Fig. 1(c). As the field increases, a sublinear dependence of $I_c$ on $B$ is observed, which is understood as follows [21]. Due to weak but finite bulk pinning, $I_c$ is determined by the bulk pinning (instead of the surfaces) when the surface barriers have been made smaller than the bulk pinning by the applied magnetic field. When this happens at a field estimated below, $I_c$ deviates from its linear dependence, the latter resulting from the edges/surfaces playing the main role. This is also the reason for the superconducting diode effect to be decreasing with increasing applied field [Fig. 1(c)]. The $I_c$ is then determined by the bulk pinning (and not the surfaces) and the inversion symmetry breaking due to surfaces starts to be irrelevant.

It is convenient to define $B_s = j_c/a$ as the field scale where $I_c$ vanishes, assuming a linear extrapolation. Our fitting parameters yield $B_s = 11.4$ Oe. Such a drastic suppression of $I_c$ by out-of-plane magnetic field in the thin film geometry could be understood by the ineffective Meissner screening [33]. The physical interpretation is that, in a thin film, the screening current is confined to the plane and is ineffective in screening the external magnetic field. As a result, the field penetrates on the scale of the Pearl length [34] $\lambda_P = 2\lambda^2/d \gg \lambda$. Maksimova [33] considered the vortex mechanism of $I_c$ and for $\lambda < W < \lambda_P$ found that $B_s = \phi_0/(\sqrt{3}\pi \ \xi W)$, where $\xi$ is the coherence length. For $W > \lambda_P$, $W$ is replaced by $2\lambda_P$ in this formula (Supplemental Material [29]). For V we estimate that $\lambda = 130$ nm and $\lambda_P = 4.25$ μm, which is somewhat smaller than $W = 8$ μm. Using $\xi = 11$ nm (Supplemental Material [29]), we obtain $B_s \sim 40$ Oe which is larger than the observed value, but is of comparable scale. This variation of $I_c$ at a field scale much smaller than the critical field and consistent with the vortex surface barrier annihilation field [20,21,33] confirms the direct role of vortices in determining $I_c$ and SC diode effect.

As the peak diode efficiency is determined by the edge asymmetry, we fabricated devices with and without a lithographically defined edge asymmetry (serrated edge with a lateral size of 3 μm, much larger than the coherence length) [35]. Critical current vs magnetic field for both devices are shown in Supplemental Material [29]. The device without defined asymmetry shows a peak value for diode efficiency of 21%, while the one with a serrated edge attains a much larger diode efficiency, reaching ~50%. Critical current peaks of the device with defined asymmetry occur at ±5.1 Oe, larger than 1.5 Oe for the device without it, which shows (and agrees with) a lower critical current for the serrated edges [35]. As temperature increases, both critical currents and diode efficiency drop as detailed in the Supplemental Material [29]. The main reason behind such a strong effect of the etching inhomogeneities is that the critical current in the devices is being determined by (i) vortex surface barrier, which is highly sensitive to the superconductor quality at the edges, and (ii) the current density at the edges. As per the contribution of (i) above, a deterioration in the superconducting properties close to an edge due to the etching-related disorder may lead to a significant lowering of the vortex surface barrier. As per the contribution (ii) listed above, the current density at the edge may be locally enhanced due to disorder for the a given total current through the superconductor [35]. These results further support the Meissner screening and asymmetric vortex surface barriers to be the underlying mechanism and a practical approach to enhance the diode efficiency [23].

Due to the highly sensitive dependence of $I_c$ on the out-of-plane field, a false "in-plane" magnetic field induced diode effect could easily be measured, with an offset between the magnetic field direction and the film plane by as small as 0.01˚ (Supplemental Material [29]). Hence, while investigating the SC diode effect under an in-plane magnetic field, any out-of-plane component of the field needs to be carefully removed to interpret the data. To study the effects of the real in-plane magnetic field on the critical currents, we developed a technique that enabled us to remove the out-of-plane field up to an accuracy of < 0.1 Oe (Supplemental Material [29]). A diode effect (type B) is then observed when the in-plane magnetic field is both parallel and perpendicular to the current flow (Supplemental Material [29]). The physical origin for this type of diode effect remains unclear and encourages further exploration – theoretical and experimental.

Experimentally, a drastic suppression of $I_c$ by an out-of-plane field has been reported in other superconductor films such as NbN [36], TaN [36], $MgB_2$ [37], (Li,Fe)OHFeSe [38] and $Nb/SrRuO_3$ bilayers [39]. However, no asymmetrical $I_c$ were reported except for an earlier work on grainy Sn films which was largely unnoticed by the community [40].

We investigated furthermore the $I_c$ rectification of the third kind (type C) in a hybrid structure where the SC film has a ferromagnetic layer EuS over it. 5nm thick EuS film shows a Curie temperature comparable to the bulk value and the hysteresis loop of EuS at 1.8K shows nice rectangular shape (Supplemental Material [29]). Fig. 2(a) shows the *I-V* scans of a Pt/V/EuS trilayer when the EuS layer was magnetized along the *y* direction (in-plane and perpendicular to the current flow). 0.2 nm of Pt is deposited to provide spin-orbit coupling and Rashba splitting at the V surface [41,42]. With a small external field to magnetize the EuS layer, a dramatic difference was observed between $I_c$ along the positive and negative directions of current flow. At $B_y$ = -30 Oe, $I_c^+$ was more than 4 times larger than $I_c^-$, producing a giant $I_c$ ratio of 480% and a diode efficiency of 65%, the highest value of diode rectification seen in superconductors yet. The $I_c$ asymmetry was reversed when the EuS magnetization direction was flipped. The temperature, magnetic field and angle dependencies of the SC diode effect were systematically studied on a second Pt/V/EuS device. A clear supercurrent rectification is also demonstrated in Supplemental Material [29]. As the temperature increased, $I_c^+$ and $I_c^-$ reduced whereas $\eta$ remained nearly unchanged from 60 mK up to 1.3 K [Fig. 2(b)]. Further increase of the temperature led to a decrease of $I_c^+$, $I_c^-$ and $\eta$, although clear diode phenomenon was seen even up to 3.6 K, close to $T_c$. At 1 K, the effect was found to quickly reach a maximum value as the field was increased to 18 Oe [Fig. 2(c)], and beyond that $I_c^+$, $I_c^-$, and $\eta$ decreased slightly as the field increased. Hysteresis in $I_c$ and $\eta$ diode efficiency was observed which resembled the magnetic hysteresis of the EuS film. This hysteresis in $I_c$ enables control of the diode polarity via the remnant EuS magnetization direction, for field-free scenario. Diode efficiency at zero external field, though a little smaller than the maximum value when EuS is fully magnetized, was still 21% for the Pt/V/EuS device. Fig. 2(d) shows $\eta$ on a polar plot: the largest asymmetry was for the magnetic field perpendicular to the current flow direction, while it was negligible when the field was parallel to the current flow.

It is tempting to interpret the giant supercurrent rectification in Pt/V/EuS as support for the finite-momentum pairing mechanism with Pt providing the required Rashba SOC [41,42] and exchange coupling with EuS giving a large spin-splitting in the V layer [43]. However, upon further investigations, we discovered that a similarly large nonreciprocity could be observed in a V/EuS bilayer device, without Pt providing the Rashba SOC [Fig. 3(a)]. Temperature [Fig. 3(b)], magnetic field and angle (Supplemental Material [29]) dependencies of the SC diode effect were systematically measured on a second V/EuS device, all showing close resemblance to that of Pt/V/EuS. Moreover, Figs. 3(c)-3(d) show similar EuS-magnetization-controlled diode effect in Nb/EuS bilayers, and persisting up to 6.5 K, as the $T_c$ for Nb was higher than V [Fig. 4(a)].

To further examine the role of the exchange field, Nb/EuS and Nb/Al$_2$O$_3$/EuS films were grown in one deposition cycle, and a mask was used to cover one of them during the 3 nm-thick Al$_2$O$_3$ film spacer layer deposition. $I$-$V$ scans at 1.8 K showed that both samples had very similar $I_c$ when EuS was magnetized [Figs. 4(b)-4(c)], $I_c$ nonreciprocity in Nb/Al$_2$O$_3$/EuS trilayer, comparable to Nb/EuS bilayer. This shows that a direct contact between the SC and FM layer was not required for the type C diode effect. Based on these observations, we conclude that neither Rashba SOC nor interfacial exchange with the FM are essential in the observed $I_c$ nonreciprocity.

The diode effect in SC/FM bilayers has been reported in other systems [31,44-48]. The phenomenon could be understood by a screening current mechanism [31] shown schematically in Fig. 4(d). The in-plane magnetization along the y axis of the FM layer produces oppositely oriented fringing fields in the $y$-$z$ plane at the two edges. A calculation shows that, for distances $r > d_m$, the fringing field can be viewed as emerging radially from point sources with opposite signs at the opposite edges and decaying as $1/r$ [49]. This produces a $z$ component of the magnetic field,

$$B_z = 2md_m z/(z^2 + y^2) \tag{5}$$

where $z$ and $y$ are measured from the sample edges and $m$ is the magnetization density. For EuS, the Eu moment is $\mu_s = 7\mu_B$ and estimate $4\pi m \sim 1.5$ T [50,51]. Similar to the type A, this perpendicular magnetic field produces a Meissner screening current flowing in the $+x$ direction, with two important differences. Since the fringing field reverses direction on the two edges, the screening current on both edges flows in the same direction [Fig. 4(d)], and a diode effect occurs without requiring edge asymmetry as in type A and the sign is independent of the material combination. This current adds to the external current in the $+x$ direction, but partly cancels it in $-x$ direction, resulting in a smaller $I_c^+$ and larger $I_c^-$. By reversing EuS magnetization, screening current around the edges flows along $-x$ direction, which reverses the current asymmetry. A second difference is that, unlike a uniform applied magnetic field, the fringing fields are strongly localized near the edges. Per our estimation (Supplemental Material [29]), the magnitude of the screening current is comparable to the departing current, leading to a gigantic diode effect.

In summary, we demonstrated ubiquitous superconducting diode effect in thin film superconductors without the need for spin-orbit or direct exchange coupling, and thus without having to invoke an exotic superconducting state harboring finite-momentum pairing. Our work shows that vortex surface barriers, and not pair breaking, determine the critical currents in two-dimensional or thin film superconductors. Hence, studying a potential finite-momentum paired superconducting order in a film using critical current nonreciprocity can only be accomplished via a careful device design that eliminates the role of vortices in determining the critical current, and as a result achieves the pair breaking mechanism. Consequently, recent reports treating the SC diode effect as a proof of finite-momentum pairing need to be reconsidered. From prospective technology development, we demonstrated a giant diode efficiency of 21% (65%) using no (small 30 Oe) external magnetic field for a nonvolatile diode effect, setting the stage for envisioning computation architectures based on superconducting rectification and providing a fresh impetus to the ongoing development of superconductors-based quantum technologies.

**Acknowledgements**

**Funding:** This work was supported by Air Force Office of Sponsored Research (FA9550-23-1-0004 DEF), Office of Naval Research (N00014-20-1-2306), National Science Foundation (NSF-DMR 1700137 and 2218550); Army Research Office (W911NF-20-2-0061, DURIP W911NF-20-1-0074). F.N. MFR, DZH acknowledge support from the European Research Council (grant number 804273). H.C. is sponsored by the Army Research Laboratory under Cooperative Agreement Number W911NF-19-2-0015. S.I. and F.S.B. are supported by European Union's Horizon 2020 Research and Innovation Framework Programme under Grant No. 800923 (SUPERTED), and the Spanish Ministerio de Ciencia e Innovacion (MICINN) through Project PID2020-114252GBI00 (SPIRIT). F.S.B. acknowledges financial support by the A. v. Humboldt Foundation. A.K. acknowledges the support by the Spanish Ministry for Science and Innovation – AEI Grant CEX2018-000805-M (through the "Maria de Maeztu" Programme for Units of Excellence in R&D). P.A.L. acknowledges the support by DOE office of Basic Sciences Grant No. DE-FG0203ER46076.

**Author contributions:** Y.H., J.S.M., L.F., F.S.B., and P.A.L. conceived and designed the study. Y.H. grew the samples and fabricated the devices. Y.H. and F.N. performed the measurements. H.C., A.L., Y.W., M.F.R., and D.Z.H. assisted with the measurements. O.G.E. and A.V. fabricated devices with defined edges and performed measurements on the devices. Y.H., A.K., P.A.L., and J.S.M. performed theory modeling. M.D., S.I., F.S.B. and F.L. provided discussion and theoretical support. Y.H., A.K., P.A.L., and J.S.M. wrote the paper with contributions from all authors.

**Competing interests**: The authors declare no competing interests.

**Data and materials availability:** The data that support the findings of this study are available from the corresponding authors upon request.

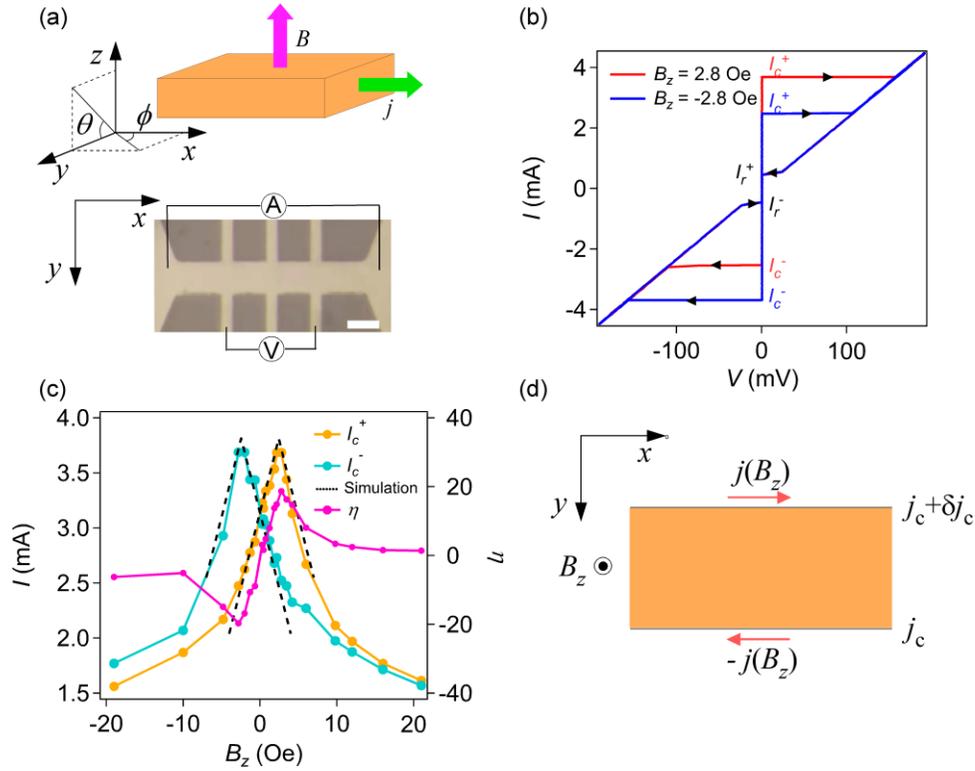

FIG. 1: Demonstration of out-of-plane field induced diode effect in a SC film at 1.8 K. (a) Top: schematic drawing of the vanadium thin film strip. Bottom: optical microscope image of the Hall bar strip of V film. Scale bar denotes 8μm. (b) *I-V* scans of the device at 2.8 Oe out-of-plane field along $\pm z$ direction as indicated by the red and blue lines. Black arrows indicate the scan direction. (c) Critical currents and diode efficiency as a function of the magnetic field. Black dashed lines show the calculated critical current values based on our model. (d) Schematic depiction of Meissner screening currents and the asymmetry between the critical current densities at the two edges.

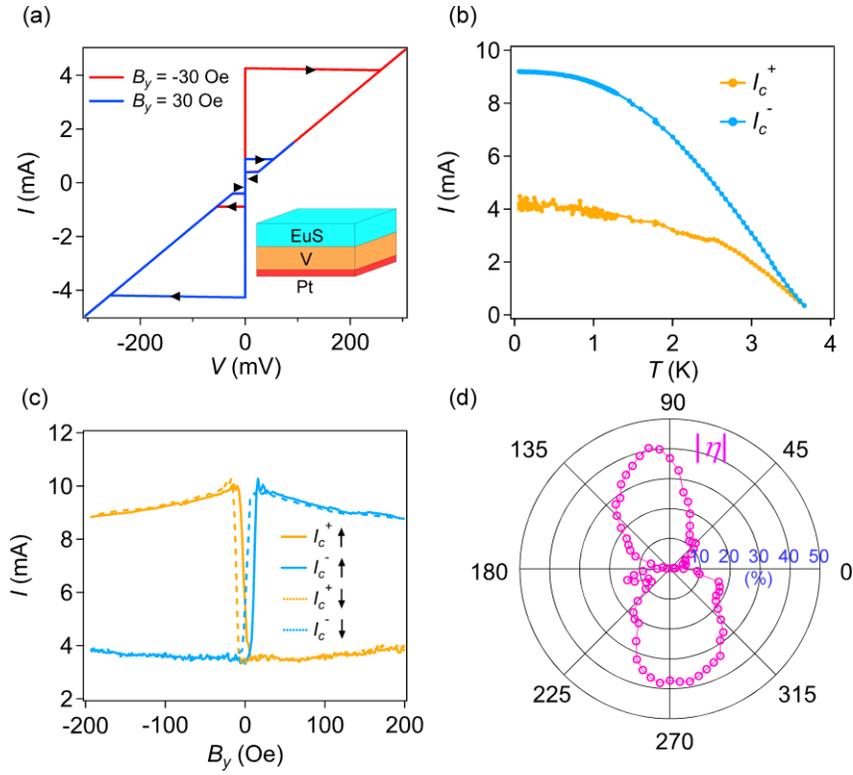

FIG. 2: SC diode effect in Pt/V/EuS trilayers. (a) *I-V* scans of a Pt/V/EuS device showing giant critical current rectification effect at 1.8 K. Inset shows a schematic of the Pt/V/EuS stack. (b) Temperature dependence of the critical current at $B_y$ = 200 Oe. (c) Magnetic field dependence of the critical current at 1 K. Solid (dashed) lines were obtained when scanning the magnetic field up (down). (d) Angle ($\phi$) dependence of $\eta$ at *T* = 1 K and *B* = 200 Oe. Data in a and b to d are from two different Pt/V/EuS devices.

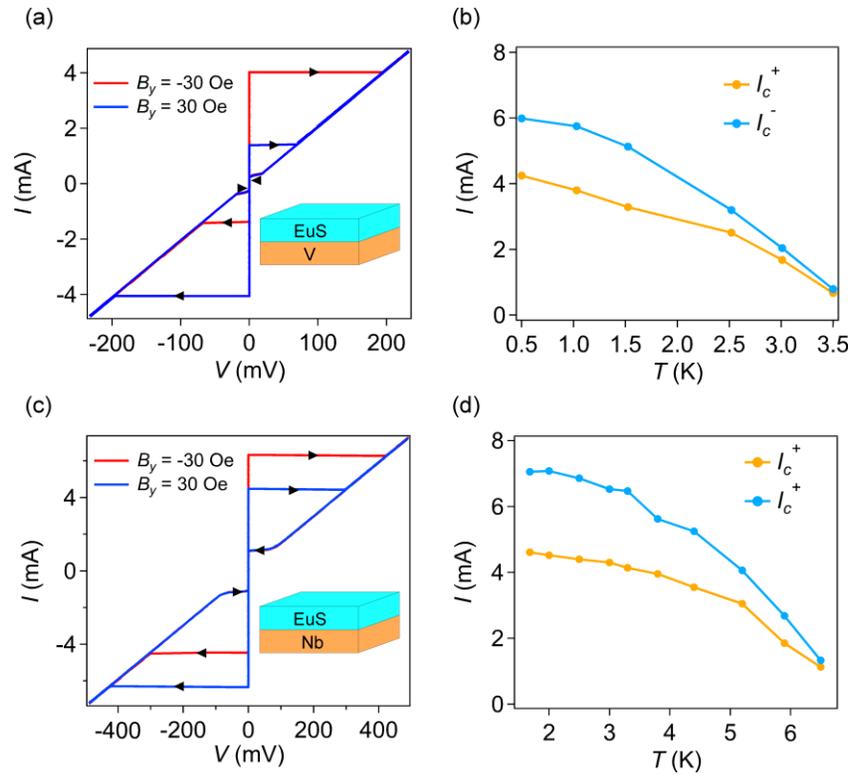

FIG. 3: SC diode effect in SC/FM bilayers. (a) *I-V* scans of a V/EuS device showing a similarly large SC diode effect as in Pt/V/EuS. Inset shows a schematic of the V/EuS stack. (b) Temperature dependence of the critical currents for a second V/EuS at $B_y$ = 30 Oe. (c) *I-V* scans of a Nb/EuS device exhibiting nonreciprocity. Inset shows a schematic of the Nb/EuS stack. (d) Temperature dependence of the critical currents at $B_y$ = 30 Oe for the same Nb/EuS device.

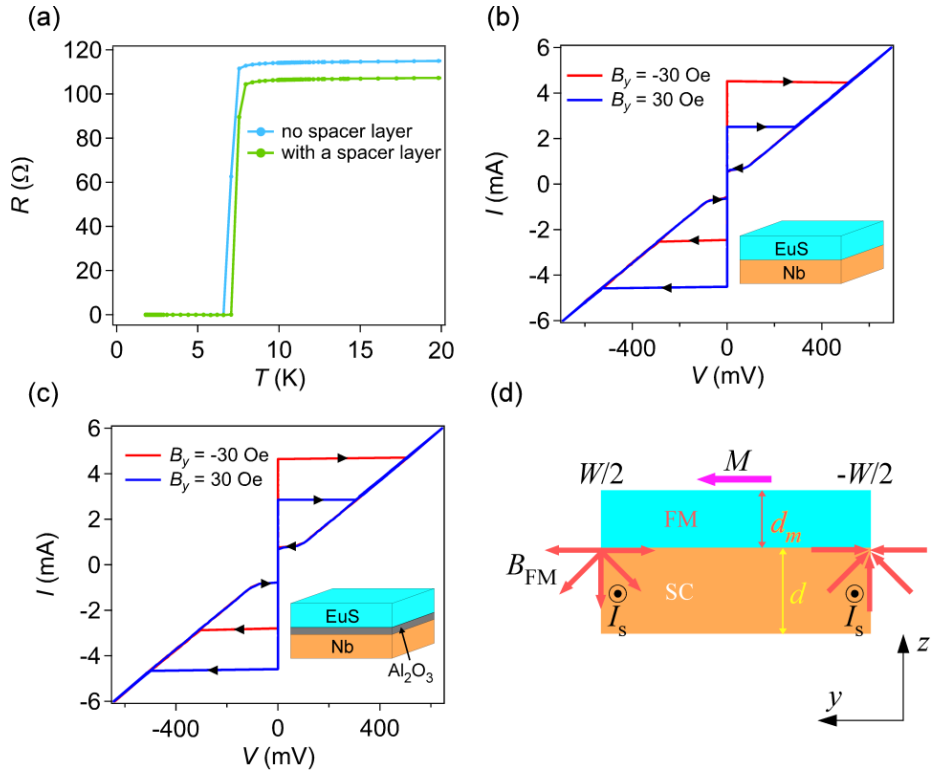

FIG. 4: Screening current mechanism for the SC diode effect in SC/FM bilayers. (a) *R-T* curves for the SC/FM films with and without a 3 nm Al$_2$O$_3$ spacer layer. (b) *I-V* scans of the Nb/EuS device measured at 1.8 K showing type C SC diode effects. Inset shows a schematic of the Nb/EuS stack. (c) *I-V* scans of the Nb/Al$_2$O$_3$/EuS device at 1.8 K showing similar diode effects as in b. Inset shows a schematic of the Nb/Al$_2$O$_3$/EuS stack. (d) Schematic depiction of the screening currents induced by the out-of-plane edge magnetic fields due to the EuS layer. *I$_s$* denotes the induced screening current.

# Supplemental material for "Ubiquitous Superconducting Diode Effect in Superconductor Thin Films"

Yasen Hou[1*], Fabrizio Nichele[2], Hang Chi[1,3], Alessandro Lodesani[1], Yingying Wu[1], Markus F. Ritter[2], Daniel Z. Haxell[2], Margarita Davydova[4], Stefan Ilić[5], Ourania Glezakou-Elbert[6], Amith Varambally[7], F. Sebastian Bergeret[5,8], Akashdeep Kamra[9*], Liang Fu[4], Patrick A. Lee[4*], Jagadeesh S. Moodera[1,4*]

* Corresponding authors. Email: yshou@mit.edu (Yasen Hou); akashdeep.kamra@uam.es (Akashdeep Kamra); palee@mit.edu (Patrick A. Lee); moodera@mit.edu (Jagadeesh S. Moodera)

**Materials and Methods**

The growth of V, Nb, EuS, Pt and $Al_2O_3$ thin films were carried out in a custom-built molecular beam epitaxy system with a base vacuum of $< 4 \times 10^{-10}$ Torr. Polished sapphire $Al_2O_3(0001)$ were used as substrates, whose surface high quality was assured by *ex situ* chemical cleaning and thermal annealing and *in situ* outgassing at 800 °C for 30 min. Subsequently the substrate was cooled down to the growth temperature for growing epitaxial metallic layers. Typical growth temperature was 230°C for V, EuS, Pt and Al2O3, and was 500°C for Nb. High-purity V, Nb, Pt and EuS sources were evaporated from an electron-beam source (e-gun) with a growth rate of approximately 0.5 Å/s. The films were patterned to a Hall-bar geometry following standard electron beam lithography and Ar ion milling.

Transport and magnetic measurements were performed in a Quantum Design Physical Property Measurement System (PPMS), equipped with a 9 T superconducting magnet. Current–voltage curves were measured through a current preamplifier (DL Instruments, model 1211), a voltage amplifier (Stanford Research Systems, model SR560) and a National Instruments data acquisition system. Ultra-low temperature measurements were realized in a dilution refrigerator equipped with a three-axis vector magnet and with a mixing chamber base temperature below 10 mK. The sample was connected via phosphor-bronze lines, which included a series of pi-filters and RC filters from QDevil, installed both at the mixing chamber level and at the sample holder. Overall, the resistance of each measurement line was about 2.6 kΩ. To minimize Joule heating, the source contact of the device (leftmost contact in Fig. 1a) was connected to 11 parallel lines, while the drain (rightmost contact in Fig. 1a) was bonded to a cold ground with negligible series resistance. In addition to this, two lateral arms of the device (two leftmost lateral contacts in Fig. 1a) were also connected in parallel to the source contact. In this way, the source-drain resistance of the entire measurement setup was 991 Ω when the device was in the superconducting state and transitioned to 1156 Ω in the resistive state.

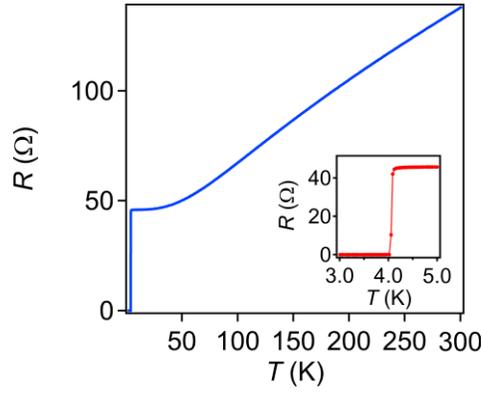

FIG. S1: *R-T* curve for a typical V device showing $T_c$ ~ 4.1K and a residual resistance ratio ~3.

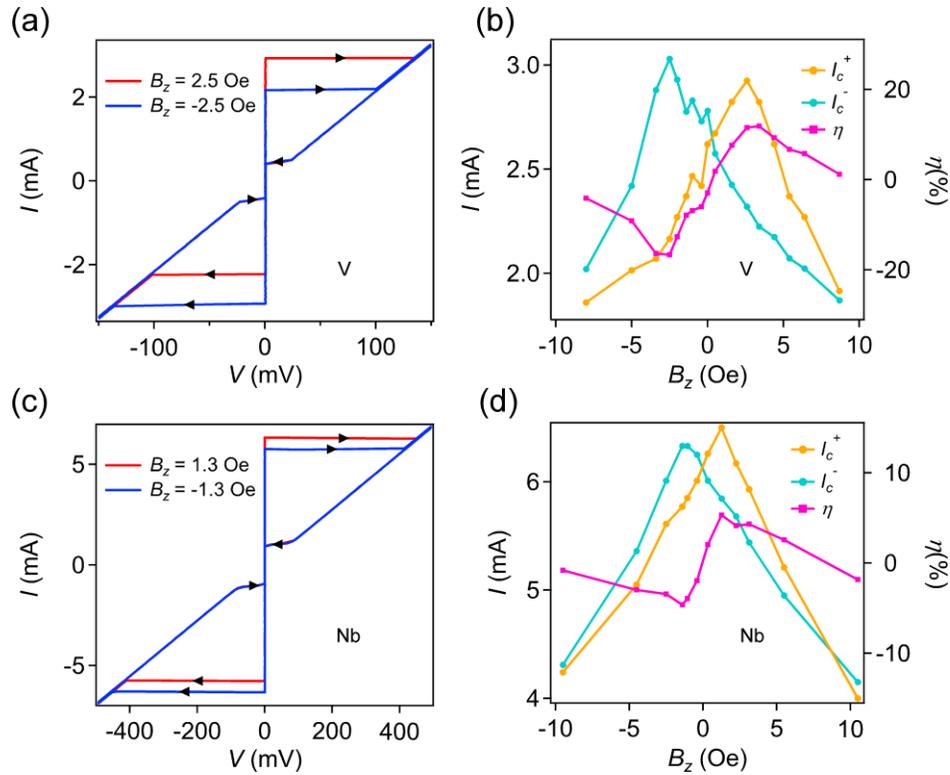

FIG. S2: Type A SC diode effect in another V film and in a Nb film at 1.8 K. (a) *I-V* scans of a second V device at 2.5 Oe out-of-plane field along $\pm z$ direction as indicated by the red and blue lines. Black arrows indicate the scan direction. (b) Critical currents and diode efficiency as a function of the magnetic field. (c and d) Similar effects in a Nb device.

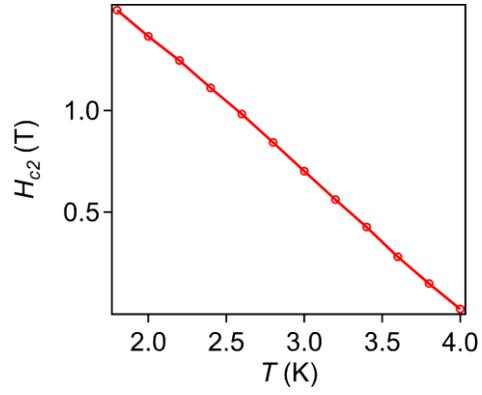

FIG. S3: Out-of-plane critical field ($H_{c2}$) vs temperature for a typical V device. The coherence length is estimated to be ~11 nm by the Ginzburg–Landau equation [52] $\varepsilon_{GL(0K)} = \sqrt[2]{\phi_0/2\pi H_{c2}(0K)}$.

**Note 1: False "in-plane" magnetic field induced diode effect.**

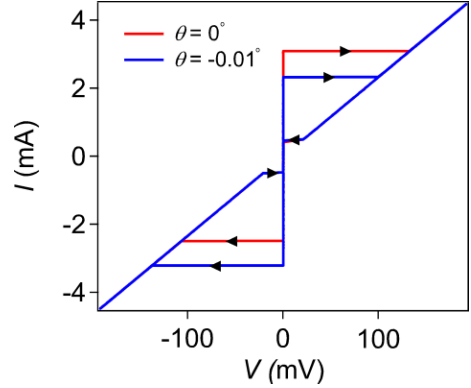

FIG. S4: *I-V* scans of the device showing a false "in-plane" diode effect, which is actually caused by the out-of-plane component of the magnetic field. Note the diode polarity is reversed when the magnetic field deviates from nominally in-plane direction by as little as 0.01˚.

In our study, the device was mounted on a rotating sample holder in a DynaCool Physical Property Measurement System (PPMS) set up, with a minimum step rotation of 0.01˚. The thin film was mounted so that the in-plane magnetic field was along the *y* axis ($\theta = 0$˚). The red curve in Fig. S4 shows the *I-V* scans of the device with a 2 T "in-plane" applied magnetic field showing different $I_c^+$ and $I_c^-$ values suggesting a SC diode effect. However, when the device was rotated to $\theta = -0.01$˚, the asymmetry of $I_c^+$ and $I_c^-$ was reversed. With such a small rotation, the change of the in-plane component of the magnetic field was ~ $3 \times 10^{-4}$ Oe, which is too small to have any effect on the device behavior. Thus, we attribute the apparent in-plane diode effect in Fig. S4 to the residual out-of-plane component of the magnetic field. Limited by the resolution of the rotator, the magnetic field was only nominally along the *y* axis and the offset angle could easily be somewhere between -0.01˚ to 0.01˚. The corresponding out-of-plane component of the magnetic field $B_z$ was up to 3.5 Oe when a 2 T field was applied to observe an apparent in-plane SC diode effect. Then the nonreciprocity polarity reversal seen is due to a sign change in the $B_z$ component. In our study, the angle between the magnetic field and the film plane was carefully calibrated so that residual offset angle was only limited by the equipment resolution. Without careful calibration, we found that the offset angle could easily be anywhere from -1° to 1°, when manually placing the device along a certain plane. In this case, the false in-plane field-induced diode effect occurs even when the applied field is as small as 100 Oe.

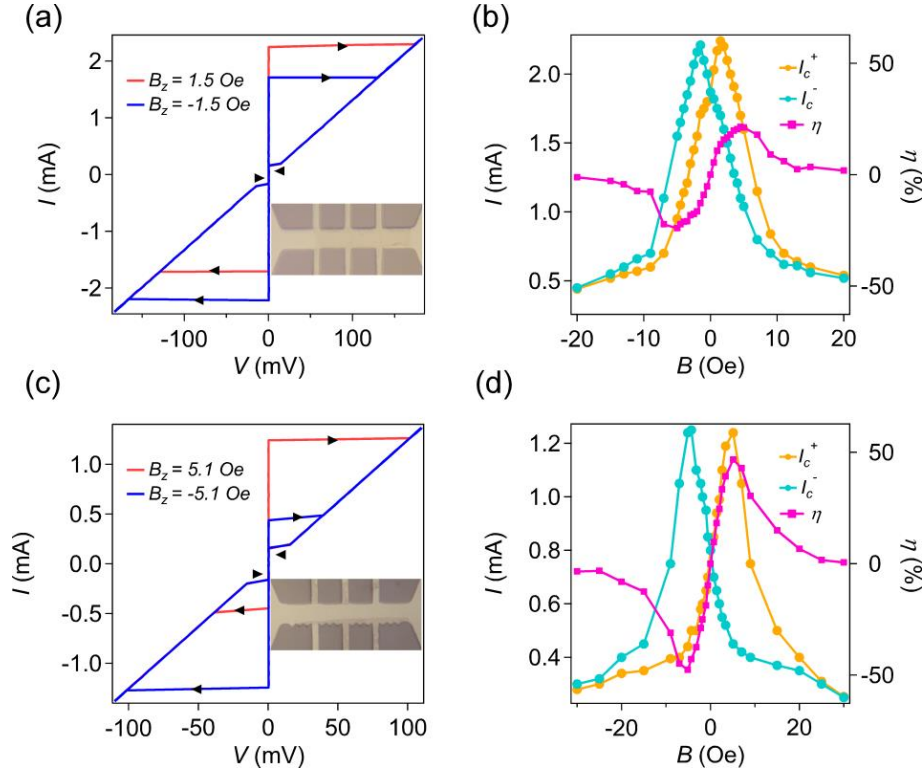

FIG. S5: Type A SC diode effect in V devices without and with defined edge asymmetry at 1.8 K. (a) $I$-$V$ scans of a V device without defined edge asymmetry at 1.5 $Oe$ out-of-plane field along $\pm z$ direction as indicated by the red and blue lines. Black arrows show the scan direction. (b) Critical currents and diode efficiency as a function of the magnetic field. (c-d) Similar effects in the V device with one serrated edge and one 'straight' edge. Insets show the device image (dark areas on the devices are PMMA residuals).

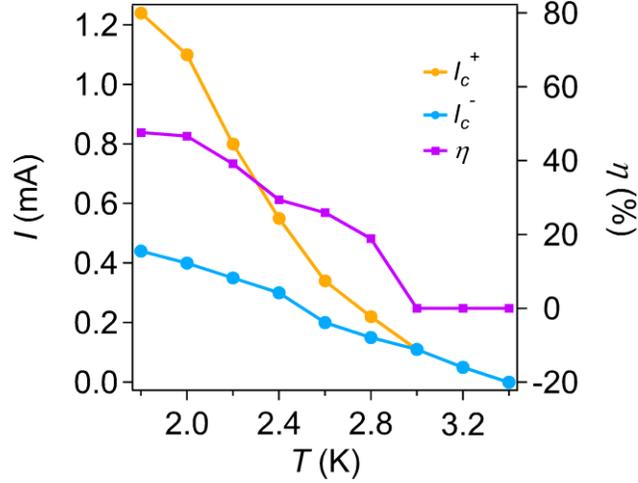

FIG. S6: Critical currents and diode efficiency of the device with defined edge asymmetry shown in Fig. S5(c) at different temperatures.

**Note 2: Suppression of critical current by out-of-plane magnetic field.**

In this section we give a physical explanation of why a very small perpendicular field can suppress the critical current by giving a simple derivation of the field $B_s$ defined in the text.

The key observation is that in a thin film, the supercurrent is confined to a plane and is ineffective in screening an external field applied perpendicular to the plane. This problem was treated by Pearl (ref. 34) who showed that the field penetrate on the scale of the Pearl length $\lambda_P = \frac{2\lambda^2}{d}$ which greatly exceeds the usual London penetration depth $\lambda$. First, we consider the case when the sample width $W \ll \lambda_P$. This was the case treated by Maksimova (ref. 33). In this case the applied field penetrates completely. In the London gauge the vector potential $A_x = B_y$ and equals $\frac{BW}{2}$ at the sample edge. We use the London equation $j_s = \frac{c}{4\pi} \frac{A_x}{\lambda^2} = \frac{c}{8\pi} \frac{BW}{\lambda^2}$. Setting $j_s$ to equal the Ginzburg-Landau critical current $j_{GL}$ given by equation (8) in the text, we find the critical field to be $\frac{2}{3} \phi_0/(\sqrt{3}\pi\, \xi W)$, which apart from the factor $\frac{2}{3}$, is the same as $B_s = \phi_0/(\sqrt{3}\pi\, \xi W)$ quoted in the main text and in ref. 33.

In the opposite case $W \gg \lambda_P$, the field is confined to a distance equal to the Pearl length near the edge. Using equation (6) the vector potential $A_x$ in the London gauge is found to be $B\lambda_P$ at the sample edge instead of $\frac{BW}{2}$ for the previous case. We can repeat the argument above but replace $\frac{W}{2}$ by $\lambda_P$. Therefore, in the expression for $B_s$ we simply replace $\frac{W}{2}$ by $\lambda_P$ in this limit.

**Note 3: Precise removal of the out-of-plane magnetic field**

To completely remove the out-of-plane component of the magnetic field, we measured the device in a dilution fridge fitted with a vector magnetic field configuration, where magnetic fields along two orthogonal directions $B_1$ and $B_2$ could be tuned individually. The device was mounted in a scheme shown in Fig. S7, where $B_1$ was along the device $y$ axis ($B_1 = B_{1y}$), with some inevitable residual angle $\theta$ that created a small perpendicular component $B_{1z}$. This residual out-of-plane component was subsequently removed by adjusting $B_2$ along the $z$ axis. Note that the offset angle $\theta$ in Fig. S7 is exaggerated for better illustration and the real angle is less than $1°$.

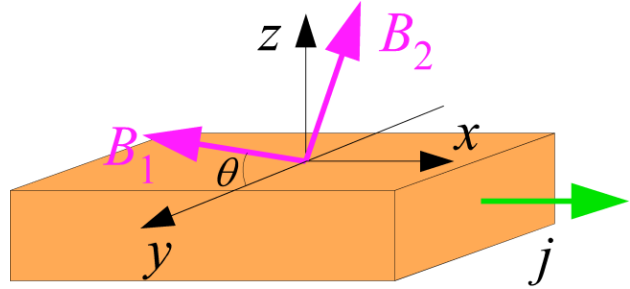

FIG. S7: Schematic depict of the device in a vector magnetic field. $B_1$ and $B_2$ are two individually adjustable magnetic fields orthogonal to each other. The device is placed so that $B_1$ is nominally in the film plane and perpendicular to the current flow direction. $\theta$ denotes the residual angle deviating from the film plane. This creates the residual out-of-plane component $B_{1z} = B_1\sin\theta$, which is then eliminated by adjusting the second magnetic field $B_2$ along the $z$ axis.

At each applied in-plane field, $B_2 = B_{2z}$ was scanned and critical currents of the device were measured. Fig. S8(a) shows $I_c^{\pm}$ as a function of $B_2$ at different $B_1$ values. We found that while $I_c^{\pm}$ vs $B_2$ always follows the inversed "V" shape, the peak position "$B_{2p}$" was shifted away from 0 when an in-plane field $B_1$ was applied. Further analysis shows that $B_{2p}$ follows $B_1$ linearly [Fig. S8(b)]. This is reasonable because when an "in-plane" field $B_1$ is applied, the out-of-plane component $B_{1z}$ suppresses the critical currents of the device. As $B_2$ is scanned along the $z$ axis, the largest critical currents are located at where $B_2$ balances out $B_{1z}$ ($B_2 = -B_{1z} = -B_1\sin\theta$).

The residual angle $\theta$ for this device was then determined to be $0.291°$ by a linear fitting. With this information, we defined a corrected out-of-plane field $B_{zc} = B_2 + B_1 \times sin0.291°$, which represents the actual out-of-plane magnetic field. Figs. S8(c)-8(e) present the critical current as a function of $B_{zc}$ with the in-plane fields $B_1$ being 0 T, 2 T and 4 T, respectively. At $B_1 = 0$ T, a typical type A SC diode effect was observed, with the diode efficiency being 0 at $B_{zc} = 0$ Oe [Fig. S8(C)]. However, when the applied in-plane field was 2 T and 4 T, a clear difference between $I_c^+$ and $I_c^-$ was observed even at $B_{zc} = 0$ Oe. Because the out-of-plane field is precisely 0, we attribute the observed supercurrent rectification at $B_{zc} = 0$ Oe to the effect of the actual in-plane magnetic field.

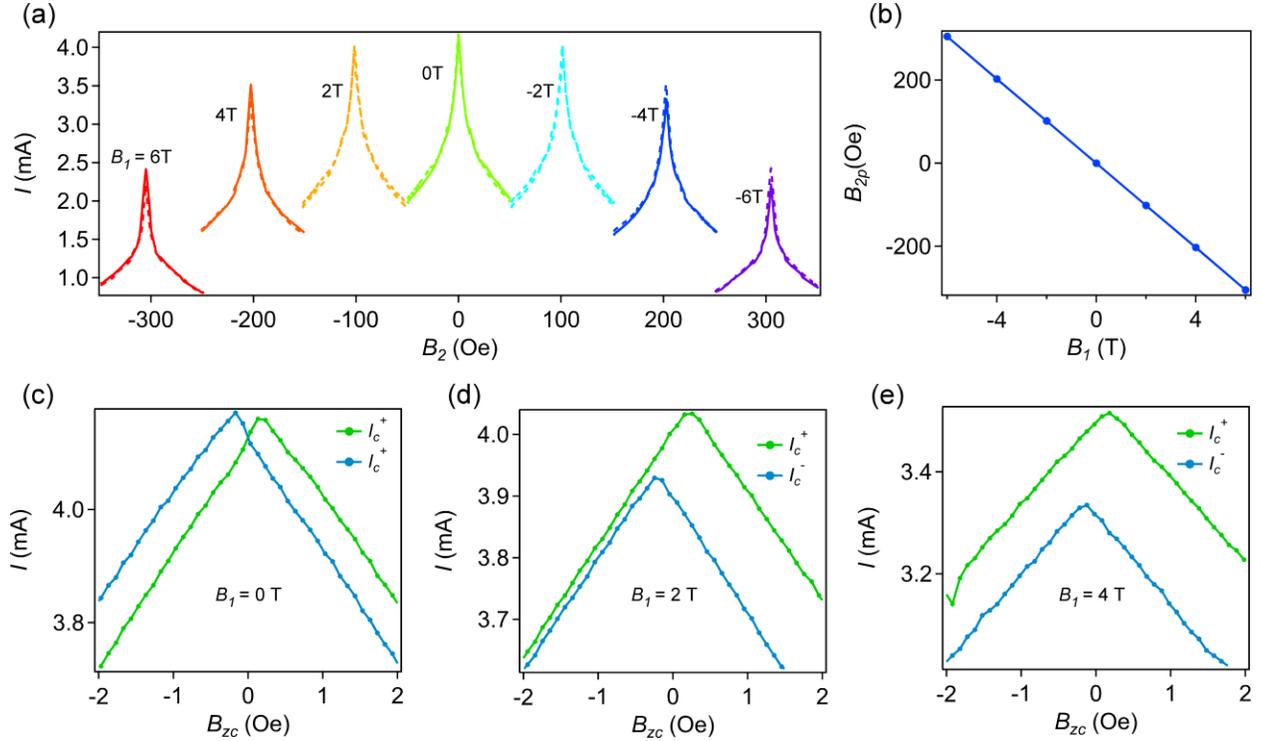

FIG. S8: Elimination of the out-of-plane magnetic field. (a) Critical currents vs $B_2$ at different values of $B_1$ for the scheme shown in Fig. S7. (b) Peak position of the critical current $B_{2p}$ as a function of the applied in-plane field $B_1$. (c-e) Critical currents vs corrected $B_{zc} = B_2 + B_1 \times sin\theta$ showing actual in-plane field induced diode effect (type B). All data were measured at $T = 500$ mK.

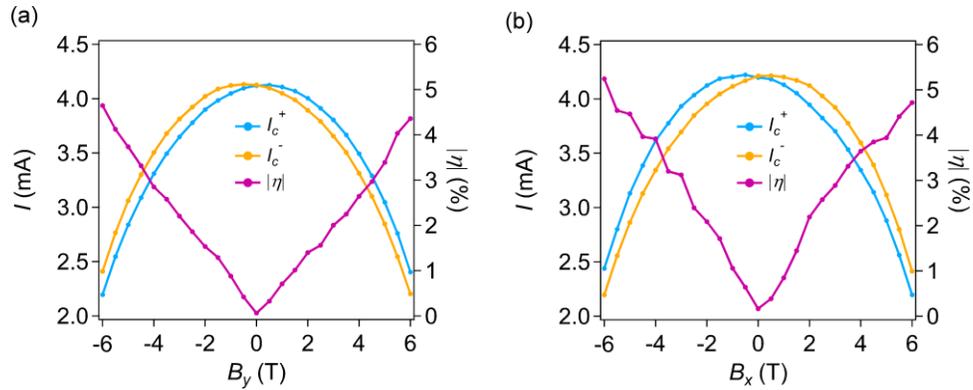

FIG. S9: In-plane field induced SC diode effect in SC V film. (a) $I_c^{\pm}$ and $|\eta|$ as a function of the in-plane field perpendicular to current flow. (b) $I_c^{\pm}$ and $|\eta|$ as a function of the in-plane field parallel to current flow. Both sets of data were taken at 500 mK.

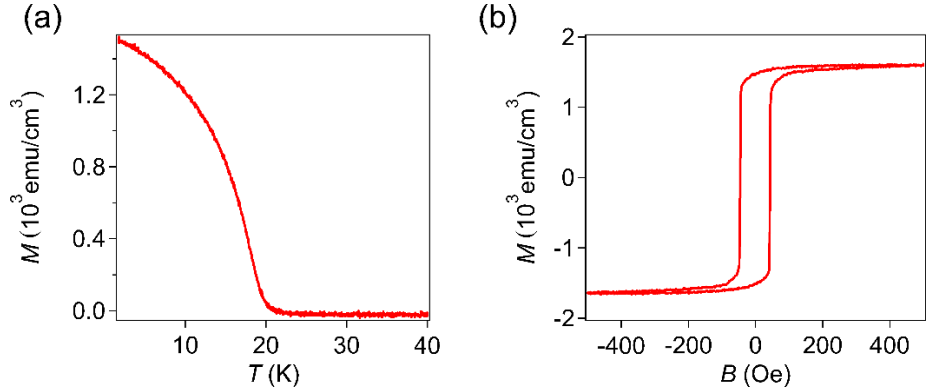

Fig. S10: Magnetic properties of a typical V(8nm)/EuS(5nm) film. (a) Measured $M$ as a function of temperature with 50 Oe in-plane applied field showing a Curie temperature of 17K. (b) Hysteresis loop at 1.8K showing a nicely rectangle loop. Linear background from the sapphire substrate and superconducting peaks are removed.

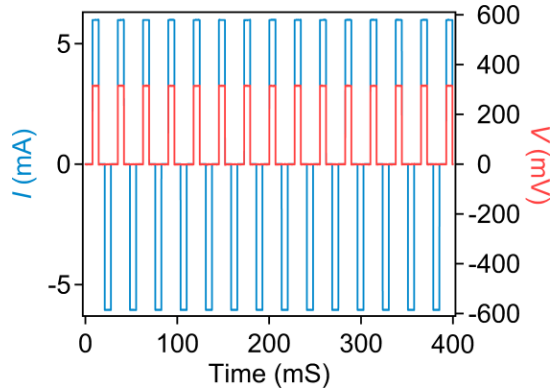

FIG. S11: Demonstration of a clear supercurrent rectification at $B_y = 200$ Oe and $T = 1$ K. The device current (blue) and the voltage drop (red) as a function of time is shown. The device behaved as a superconductor supporting dissipationless current along the -$x$ direction, while dissipative current flows through a normal metal in the other direction.

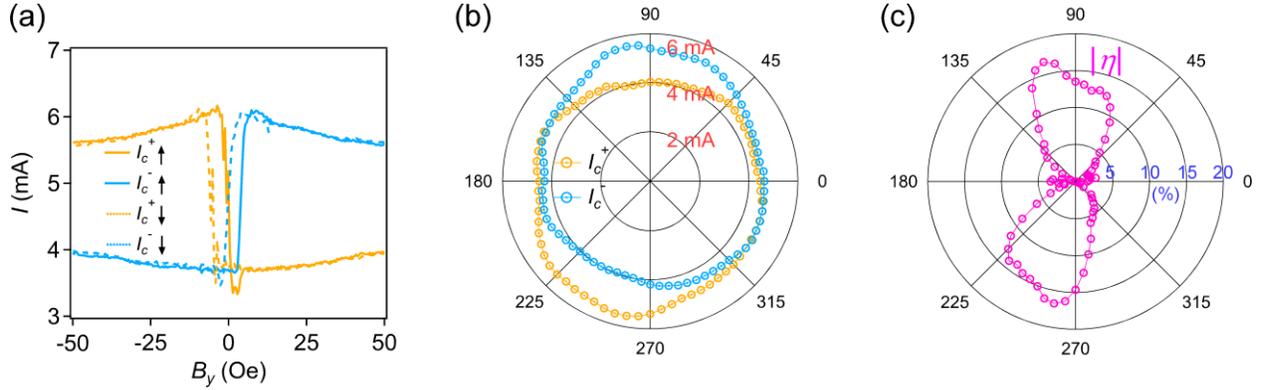

FIG. S12: Magnetic field and angle dependence of the diode effect for the second V/EuS device. (a) Magnetic field dependence of the critical currents at 1K. Solid (dashed) lines were obtained when scanning the magnetic field up (down). (b) Angle ($\phi$) dependence of $I_c^{\pm}$ at $T = 1$ K and $B_y = 50$ Oe. (c) The corresponding diode efficiency as a function of angle $\phi$.

**Note 4: Estimation of the screening current in FM/SC bilayers.**

The magnitude of the screening current is estimated using the London equation. The $x$ component of the vector potential is obtained by integrating $\boldsymbol{B} = \nabla \times \boldsymbol{A}$, i.e., $B_z = \frac{\partial A_x}{\partial y}$. In the London gauge we find:

$$A_x(y) = \int_0^y B_z(z, y')\, dy' \tag{S1}$$

where we have shifted the $y$ coordinate so that the sample lies between $y = -W/2$ and $W/2$. Using Eq. (5) in the manuscript in the shifted coordinate, we find that right at the sample edge, the $A_x$ field is independent of $z$ and is given by $A_x = \pi m d_m$. From the edge it extends into the sample by a distance $\sim d$ in the $y$ direction. Using the London equation $j_x = \frac{c}{4\pi}\frac{A_x}{\lambda^2}$, we find that, at the sample edge, the current is given by:

$$j_0 = \frac{c}{4\pi}\frac{\pi m d_m}{\lambda^2} \tag{S2}$$

On the other hand, the critical current density of the SC film is estimated by the GL theory as:

$$j_{GL} = \frac{c}{4\pi}\frac{\phi_0}{3\sqrt{3}\pi \xi \lambda^2} \tag{S3}$$

The ratio between the screening current density and the critical current density is then:

$$\frac{j_0}{j_{GL}} = \frac{3\sqrt{3}\pi^2 m d_m \xi(T)}{\phi_0} \qquad (S4)$$

By using $4\pi m = 1.5T, d_m = 5nm, \xi = 11nm, \phi_0 = 2 \times 10^3 T(nm)^2$, we estimate:

$$\frac{j_0}{j_{GL}} = 0.156 \qquad (S5)$$

This rough estimate shows that the screening current due to the fringing field is comparable to the critical current and can lead to the diode effect. The ratio is proportional to the thickness of the FM film and can be increased using thicker films, keeping in mind that the estimate is valid only for $d_m < d$. In understanding our observed effect, we experimentally find that vortex surface barriers and Meissner response of the applied magnetic field will determine the critical current in wide films. This barrier, and consequently the critical current, scales inversely with the device width. Hence, for narrow devices with widths comparable to or smaller than the coherence length, the surface barrier becomes large and the role of Meissner currents should become negligible. Under those conditions, the depairing mechanism may determine the critical current and finite-momentum Cooper pairing can be probed. The thickness of the devices should be kept small, ideally comparable to or smaller than the coherence length.